\documentclass[prl,aps,twocolumn,showpacs,preprintnumbers,amsmath,amssymb]{revtex4}

\usepackage{graphicx}
\usepackage{dcolumn}
\usepackage{color}

\begin{document}

\def\EF{$E_\textrm{F}$}
\def\cred{\color{red}}
\def\cblue{\color{blue}}
\definecolor{dkgreen}{rgb}{0.31,0.49,0.16}
\def\cgreen{\color{dkgreen}}

\title{
Electronic structure of YbB$_{6}$:
Is it a Topological Insulator or not? }

\author{Chang-Jong Kang$^1$}
\author{J. D. Denlinger$^2$}
\email[]{jddenlinger@lbl.gov}
\author{J. W. Allen$^3$}
\author{Chul-Hee Min$^4$}
\author{F. Reinert$^4$}
\author{B. Y. Kang$^5$}
\author{B. K. Cho$^5$}
\author{J.-S. Kang$^6$}
\author{J. H. Shim$^{1,7}$}
\author{B. I. Min$^1$}
\email[]{bimin@postech.ac.kr}
\affiliation{
$^1$Department of Physics, PCTP,
	Pohang University of Science and Technology, (POSTECH)
	Pohang 37673, Korea \\
$^2$Advanced Light Source, Lawrence Berkeley Laboratory,
	Berkeley, CA 94720, U.S.A.\\
$^3$Department of Physics, Randall Laboratory,
	University of Michigan, Ann Arbor, MI 48109, U.S.A. \\
$^4$Universit\"{a}t W\"{u}rzburg, Experimentelle Physik VII,
	97074 W\"{u}rzburg, Germany \\
$^5$School of Materials Science and Engineering, GIST,
	Gwangju 61005, Korea \\
$^6$Department of Physics, The Catholic University of Korea,
	Bucheon 14662, Korea\\
$^7$Department of Chemistry and Division of Advanced Nuclear Engineering, POSTECH,
	Pohang 37673, Korea
}
\date{\today}

\begin{abstract}
To finally resolve the controversial issue of whether or not the electronic structure of
YbB$_{6}$ is nontrivially topological, we have made a combined study using
angle-resolved photoemission spectroscopy (ARPES)
of the non-polar (110) surface and density functional theory (DFT).
The flat-band conditions of the (110) ARPES avoid the strong band bending effects of the polar (001) surface
and definitively show that YbB$_{6}$ has a topologically trivial B 2$p$-Yb 5$d$ semiconductor band gap of $\sim$ 0.3 eV.
Accurate determination of the low energy band topology in DFT requires
the use of a modified Becke-Johnson exchange potential incorporating spin-orbit
coupling and an on-site Yb $4f$ Coulomb interaction $U$ as large as 7 eV.
The DFT result, confirmed by a more precise GW band calculation,
is similarly that of small gap non-Kondo non-topological semiconductor.
Additionally the pressure-dependent electronic structure of YbB$_{6}$
is investigated theoretically and found to transform into a $p$-$d$ overlap
$semimetal$ with small Yb mixed valency.
\end{abstract}

\pacs{71.20.Eh,71.27.+a,75.30.Mb}


\maketitle

A great deal of recent attention has been paid to
the topological nature of
strongly correlated systems,
which include the topological Mott insulator \cite{Raghu08,Pesin10},
the fractional topological insulator \cite{Sheng11,Maciejko15},
and the topological Kondo insulator (TKI) \cite{Dzero10}.
In these systems, the interplay between topological characteristics
and strong electron correlations provides new interesting phenomena
that can possibly be utilized
for spintronic and quantum computing applications.

The first candidate material for a TKI is SmB$_{6}$,
which has been predicted first
theoretically \cite{Dzero10,Takimoto11,Dzero12,Lu13},
and then studied intensively
by transport \cite{Wolgast13,Kim13,Kim14},
angle-resolved photoemission spectroscopy
(ARPES) \cite{Denlinger13,Neupane13,Xu13,Min14},
and scanning tunneling microscopy/spectroscopy
(STM/STS) \cite{Yee13,Rossler14} experiments to explore its surface states.
Subsequently, other $4f$-electron systems have been proposed
as TKI's and topological Kondo semimetals
\cite{Yan12,Zhang12,Deng13,Li14,Weng14,Kang15,Kasinathan15}.
Two essential common ingredients for a non-trivial topological character are
(i) band inversion between opposite parity 4$f$ and 5$d$ states,
caused by rare-earth mixed-valence,
and (ii) a large spin-orbit coupling (SOC) provided by the 4$f$ states.
At the simplest level a strongly correlated bulk topological insulator (TI)
would have the generic TI property of protected, symmetry-required,
spin-textured metallic Dirac cone surface states
that span the insulating bulk gap.

YbB$_{6}$ of our present interest was proposed to be a TKI
with the mixed-valence state of Yb being 2.2 ($n_{f}$ = 13.8)
based on the inverted Yb $4f$-$5d$ bands obtained
in the density-functional theory (DFT) + Gutzwiller band
method \cite{Weng14}.
However, early photoemission \cite{Kakizaki93}
and recent ARPES \cite{Xia14,Xu14,Frantzeskakis14,Neupane15}
show that the binding energy (BE) of the Yb $4f_{7/2}$ band
is about 1 eV, indicating that there would be no $f$-$d$
band inversion and so YbB$_{6}$ would not be a TKI.
Then, inspired by the observation of (001) surface states
having the appearance of Dirac cones \cite{Xia14,Xu14,Neupane15},
two ARPES groups proposed that YbB$_{6}$ would be a weakly
correlated TI with band inversion between opposite
parity Yb $5d$ and B $2p$ bands \cite{Xu14,Neupane15}.
The topological origin of the observed surface states was questioned
\cite{Frantzeskakis14}, however, because they were observed to
not follow the expected linear Dirac cone dispersion and to exhibit time-dependent
changes. Instead band bending and surface quantum well confinement
arising from the (001) polar surface was suggested,
while not explicitly proposing that YbB$_{6}$ is not a TI.



The $p$-$d$ band inversion TI scenario was supported theoretically with
DFT + SOC + $U$ ($U$ = 4 eV) calculations \cite{Neupane15,Chang15},
but also with an incorrect 0.3 eV BE of the Yb $4f_{7/2}$ state and
in contradiction to an earlier calculation \cite{Jun07} using $U$ = 7 eV
that obtained a $p$-$d$ inverted $semimetal$ with the correct experimental Yb 4$f$ energy.
These current experimental and theoretical uncertainties
have prevented a consensus on the topological nature of YbB$_{6}$.

In this Letter, we report new ARPES experiments that definitively demonstrate
the non-Kondo non-TI electronic structure of YbB$_{6}$
and new DFT theory that agrees well with the experimental results
and strongly supports the same conclusion.
ARPES for the \emph{non-polar} (110)
surface reveals a clear $p$-$d$ semiconductor gap with no in-gap surface states,
whereas all surfaces of a TI system must have surface states.
Calculations incorporating the SOC and $U$ into the modified Becke-Johnson (mBJ)
potential \cite{TB09} describe properly the BE of the Yb $4f_{7/2}$ band
and the observed ARPES spectra of a topologically trivial Yb $5d$-B $2p$ band gap.
We have also investigated the pressure-dependent electronic structure of YbB$_{6}$
and found that the high pressure phase is a topologically non-trivial $p$-$d$ overlap
$semimetal$ with an Yb $4f_{7/2}$ BE of $\sim 0.5$ eV,
rather than an full insulator.
This result explains a recent experimental study of transport
and Yb valence under pressure \cite{Zhou15}.

ARPES measurements were performed at the MERLIN Beamline 4.0.3
at the Advanced Light Source in the photon energy ($h\nu$) range of 30$-$150 eV.
An elliptically polarized undulator was employed,
which allows selection of $s$- and $p$-polarization of the incident light.
A Scienta R8000 hemispherical electron energy analyzer was used
with energy resolution set to $\approx$ 20 meV \cite{suppl}.
Measurements were performed in a vacuum of better than
$5 \times 10^{-11}$ Torr
for the sample cooled down to $\approx$ 30 K.

The band calculations were performed using
the full-potential linearized augmented plane-wave (FLAPW)
band method, as implemented in the WIEN2K package \cite{wien2k}.
For the DFT calculations, the PBE (Perdew-Burke-Ernzerhof) exchange-correlation
functional was used in the GGA (generalized-gradient approximation).
In the GGA + SOC + $U$ method, a correlation energy of $U=7$ eV was
chosen to obtain the correct experimental value of the
Yb 4\emph{f} BE of $\approx$ 1 eV,
which is consistent with the previous calculations \cite{Jun07,Kunes04}.
The mBJ potential is adopted to provide band gap corrections
in agreement with the improved many-body but
much more computation-demanding GW calculation \cite{TB09,Singh10}.
The details of the calculational methods are described
in the Supplement \cite{suppl}.

\begin{figure}[b]
\includegraphics[width=8.5 cm]{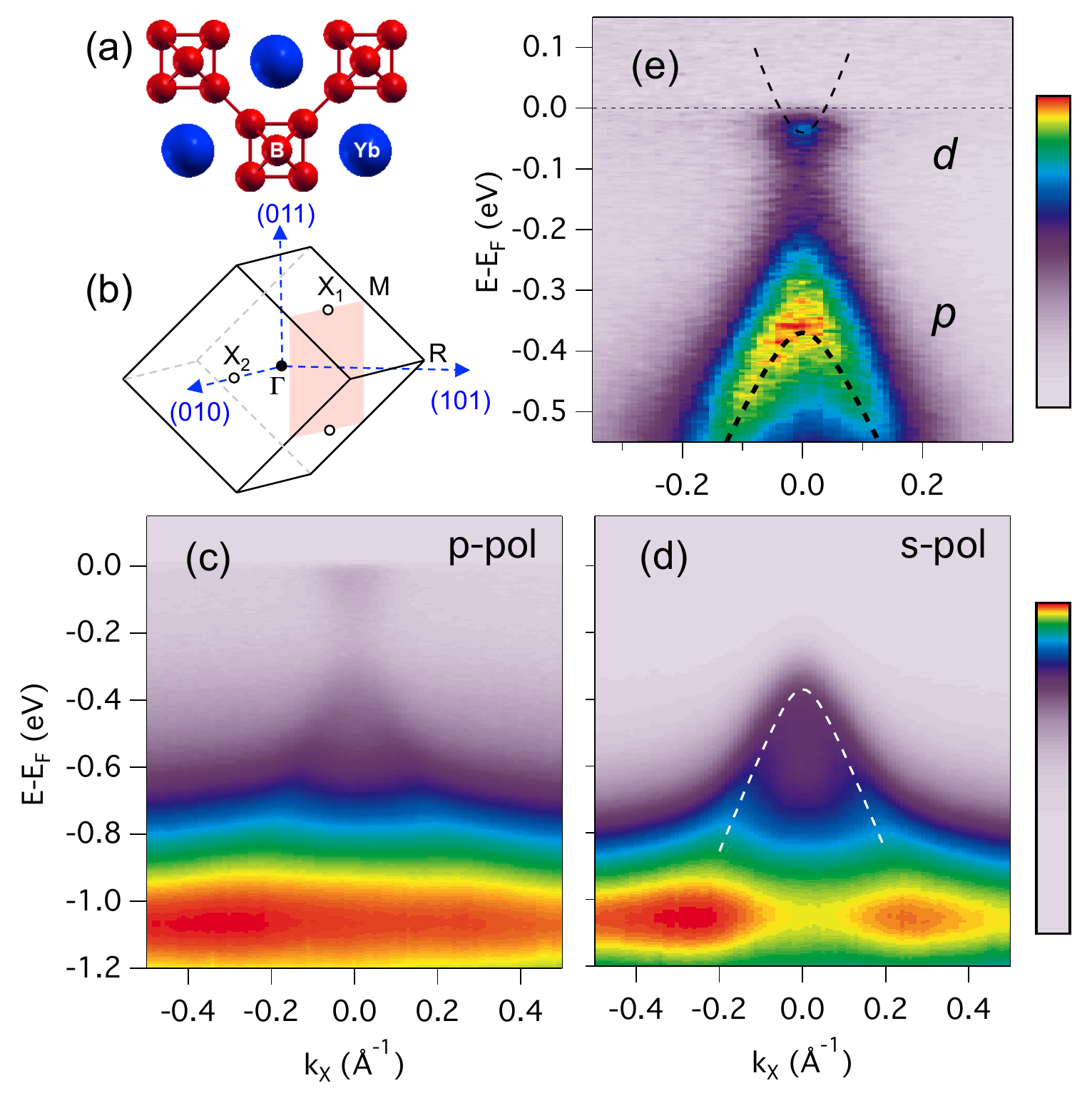}
\caption{(Color Online)
(a) Schematic structure of the non-polar YbB$_{6}$ (110) surface.
(b) Cubic BZ with (110) orientation illustrating the locations of bulk X-points.
(c,d) X-point spectra measured at  $h\nu$ = 120 eV with $p$- and $s$-polarization,
illustrating the opposite polarization dependence of $p$-hole and $d$-electron states
(e) Zoom of the $p$-polarization spectrum with the Yb 4$f$ spectral intensity removed
to enhance the view of the $\sim$0.3 eV band gap.
Dashed lines are non-parabolic fits to the spectral intensity maxima (see text).
}
\label{arpes}
\end{figure}

\begin{figure}[t]
\includegraphics[width=8.5 cm]{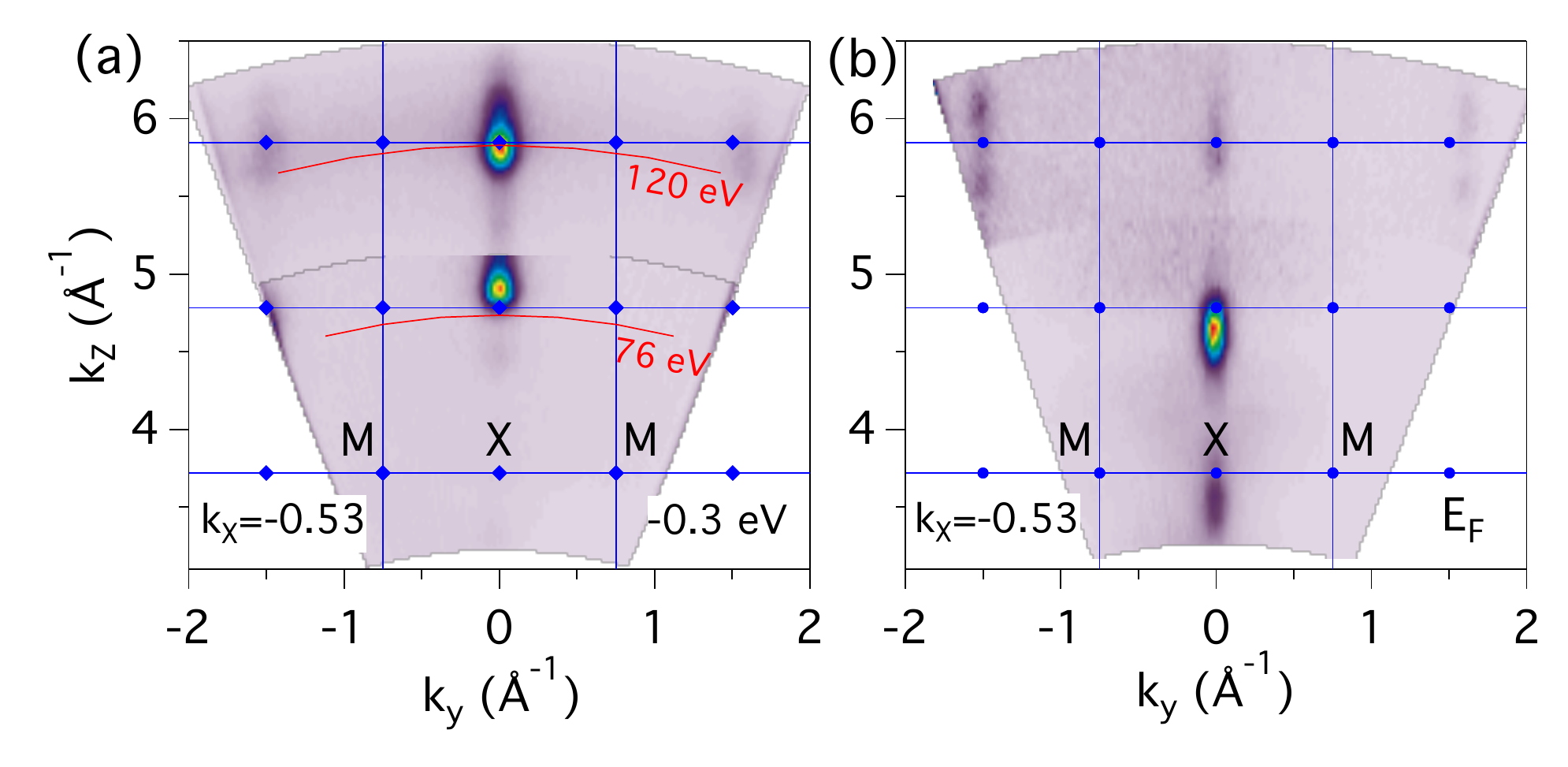}
\caption{(Color Online)
Off-normal photon-dependent map of YbB$_{6}$ (110)
at fixed $k_{x}$ = $-0.53$ ${\AA}^{-1}$, as shown in Fig. \ref{arpes}(b),
of the spectral intensity of (a) valence band at $-0.3$ eV
and (b) conduction band at \EF.
The 3D bulk-like $k_{z}$ dependences of both the valence and conduction bands
confirms the flat band conditions of the non-polar (110) surface.
}
\label{photon}
\end{figure}

For insulating hexaborides, the polarity of (001) surfaces
with different charge terminations
can lead to $n$- and $p$-type band bending and quantum well states
that make it difficult for ARPES to directly observe the bulk band gap.
Also spectra from spatially inhomogeneous regions ({\it i.e.},
both $n$- and $p$-type) can falsely appear
to show Dirac cones or $p$-$d$
overlap \cite{Frantzeskakis14,Denlinger15}.
While surface modification and aging provide some control over
the band bending and assist in the ARPES
interpretations \cite{Denlinger15},
these problematic band-bending effects can be avoided
by instead measuring a non-polar surface such as the (110) surface
whose charge neutral bulk-termination is schematically illustrated
in Fig. \ref{arpes}(a).
For this purpose, a (110) surface of YbB$_{6}$ was prepared
from the natural facet of a single crystal grown by the
aluminum-flux method. After etching in hydrochloric
acid and ion sputtering of the surface, the sample was
annealed to 1300 $^{\circ}$C in ultra high vacuum to produce a
spatially uniform 1 $\times$ 1 ordered surface \cite{suppl}.

X-point ARPES spectra measured along M-X-M at $h\nu$ = 120 eV
using two different linear polarizations of the incident light is shown
in Figs. \ref{arpes}(c) and (d).
Above a strong Yb 4$f$ peak at $-1.05$ eV, the $p$-polarization spectrum shows
a weak hole-band dispersion and a small electron-like intensity at \EF.
A strong polarization selectivity of these states is revealed
by the $s$-polarization spectrum in Fig. \ref{arpes}(d),
where the electron conduction state is totally suppressed
and the valence hole band is strongly enhanced
to manifest a triangular-like dispersion
with a rounded-maximum and hybridization interaction with the Yb 4$f$ states.
The strong hole-band intensity allows a quantitative fit (dashed line)
to a two-band $k\cdot p$ non-parabolic dispersion model \cite{suppl}
with a band maximum of 0.35 eV below \EF.

Figure \ref{arpes}(e) shows an enlarged view of the $p$-polarization spectrum
in which the Yb 4$f$ spectral intensity tail has been divided out to obtain an enhanced
image of the $\sim$ 0.3 eV semiconductor band gap
between the B 2$p$ valence and Yb 5$d$ conduction states.
To further characterize the conduction band dispersion and energy minimum,
K-dosing of the surface was used to induce a small $n$-type band bending
until the electron pocket was increased in depth to 0.2 eV
revealing enough of a dispersion \cite{suppl}
to perform similar non-parabolic dispersion analysis.
The resulting process exhibited no discernible surface band gap narrowing,
thus allowing evaluation of a band gap of 0.32 eV.

\begin{figure}[t]
\includegraphics[width=7.5 cm]{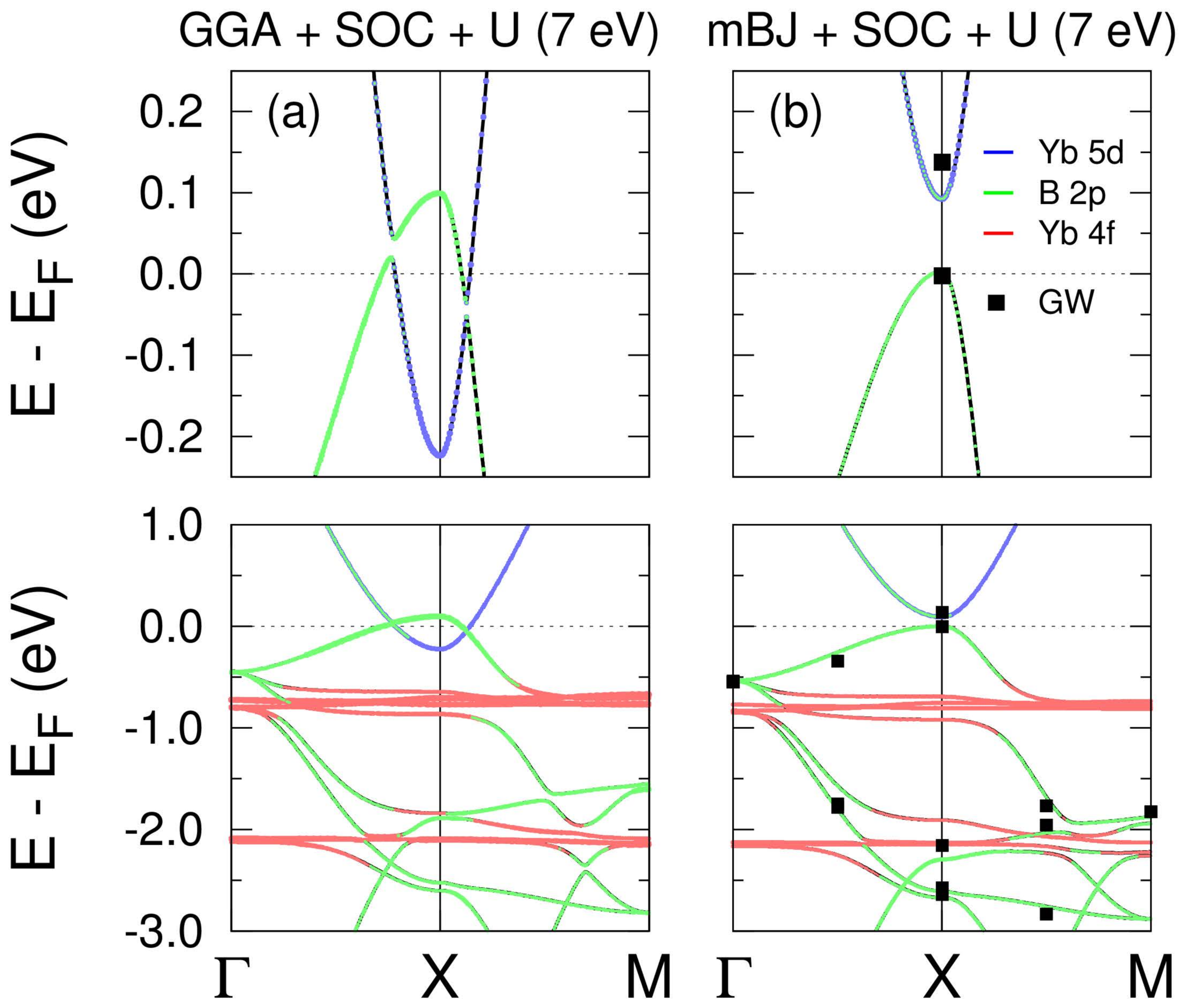}
\caption{(Color Online)
DFT-Wien2k band structures of YbB$_{6}$.
(a) GGA + SOC + U (7 eV) calculation yields a semimetallic $p$-$d$ overlap with anti-crossing gaps.
(b) mBJ + SOC + U (7 eV) bands overlaid with open-core GW band results (dots).
Both exhibit semiconductor band gaps.
}
\label{dft}
\end{figure}

To explicitly confirm that the bands shown in Fig. \ref{arpes} are bulk,
we have measured their $k_{z}$ dependences in the process of locating the bulk X-points.
Figures \ref{photon}(a) and (b) show the $k_{y}$-$k_{z}$ maps
at fixed $k_{x} = -0.53 {\AA}^{-1}$ for the valence band
at $-0.3$ eV and the conduction band at \EF, respectively.
Both constant energy cuts
exhibit strong intensity features close to bulk X-points
at $k_{y}$ = 0 for $h\nu$ = 76 eV and 120 eV
as well as intensities at X-points of the second Brillouin zone (BZ)
at $k_{y} = \pm1.5 {\AA}^{-1}$.
The small vertical $k_z$-elongation of the X-point intensities in Fig. \ref{photon}
is well accounted for by the inherent bulk band structure anisotropy
(see Fig. \ref{dft}) and
the $k_z$-broadening effect resulting from the finite inelastic mean free path
of the photoelectrons.
The pinning of \EF\ at the bottom of the conduction band
is consistent with the negative sign of the bulk Hall coefficient
\cite{Tarascon80,Kim07,Zhou15}, and consistent with flat-band conditions
of the non-polar (110) surface.
Hence both the valence and conduction bands shown in Fig. \ref{arpes}
are 3D-like bulk bands
and do not originate from the 2D-like surface states.
The strong polarization dependence in Fig. \ref{arpes}(d) also independently
confirms that these states are not linear Dirac cone dispersions,
which would instead exhibit some continuity of the same orbital characters
between the upper and lower parts of the Dirac cone.

The bulk X-point spectrum in Fig. \ref{arpes}(e) exhibiting a clear
small direct semiconductor gap between  valence and conduction band states
and the absence of in-gap surface states
is the central experimental result of this study.
The (110) ARPES definitively proves the absence of a $p$-$d$ overlapping band structure
and hence a lack of parity inversion that is the key first requirement
for a topological electronic structure interpretation of previous ARPES for the (001) surface.
Therefore, the observed chirality in 2D surface states of YbB$_{6}$ (001) in
circular-dichroism (CD) \cite{Xia14} and spin-resolved ARPES \cite{Xu14},
cited to support the TI scenario of single-spin in-gap states,
must have alternative explanations.  Geometrical and final state effects
are known to allow the detection of CD and spin-polarization
in photoemission of non-chiral and non-magnetic solids \cite{Schohense90,sarpes},
and can prevent an unambiguous detection of spin-polarization asymmetries
in YbB$_{6}$, as discussed elsewhere \cite{Denlinger15}.




Next we turn our attention to theoretical predictions of the YbB$_{6}$
electronic structure using the DFT method.
We first reproduce the literature result \cite{Jun07} of a GGA + SOC + $U$ (7 eV)
calculation for YbB$_{6}$ in Fig. \ref{dft}(a),
which predicts a semimetallic band structure
with a $p$-$d$ band overlap at \EF.
The local gapping at the band crossing points
arises from rather weak 5$d$ SOC \cite{Cross}.
Since the $p$-$d$ overlap anti-crossing points vary
in energy around the X-point,
the small local gapping cannot produce a full bulk gap,
resulting in a complex semimetallic Fermi surface (FS),
consisting of ``lens'' hole and ``napkin ring'' electron sheets.
The calculated YbB$_{6}$ 4$f$ BE of 0.7-0.8 eV
relative to the valence band maximum is in agreement with the
experimental ARPES result in Fig. \ref{arpes} of 1.05 eV which
includes the 0.32 eV band gap.
The location of the 4$f$ state far from \EF\ results in
only a minor influence on the semimetallic FS that is thus very similar
to predictions of the non-rare-earth divalent hexaborides \cite{Massidda00,Hasegawa79,Rodriguez00}.

In Fig. \ref{dft}(b), we present an mBJ + SOC + $U$ (7 eV) band result,
overlaid with open-core pseudopotential single pass GW band result (dots) \cite{suppl}.
In both cases, the small $p$-$d$ overlap of the GGA + SOC + $U$
calculation in Fig. \ref{dft}(a)
is transformed into a small $\approx$ 0.1 eV semiconductor gap
with good quantitative agreement between the two methods \cite{suppl}.
This result clearly indicates that YbB$_{6}$ is
a topologically trivial small band-gap semiconductor.
Not surprisingly, slab calculations for both the YbB$_{6}$ (001) and (110) surfaces
also show no topological in-gap surface states (see the Supplement \cite{suppl}).

This semiconductor result is reminiscent of the case of CaB$_{6}$,
whose early DFT-based semimetallic model for anomalous transport
was revised to be that of a 1 eV semiconductor
with the assistance of GW theory \cite{Tromp01},
and subsequently confirmed with ARPES \cite{Denlinger02}
and other experiments using high-purity boron samples \cite{Cho04,Rhyee04}.
This straightforward theoretical prediction for YbB$_{6}$
of being a topologically trivial semiconductor
is in contrast to two recent calculations that predict YbB$_{6}$
to be a TI based on $f$-$d$ band inversion \cite{Weng14}
or $p$-$d$ inversion \cite{Neupane15,Chang15}.
The flaws in these previous band calculations, resulting in incorrect
Yb 4$f$ binding energies and mixed valency, are discussed in detail
in the Supplement \cite{suppl},
along with
angle-integrated valence band spectra from the (110) surface that provide
definitive proof of the pure Yb divalency in YbB$_6$ \cite{suppl},
and thus additionally rule out these erroneous theory calculations.

\begin{figure}[t]
\includegraphics[width=8.5 cm]{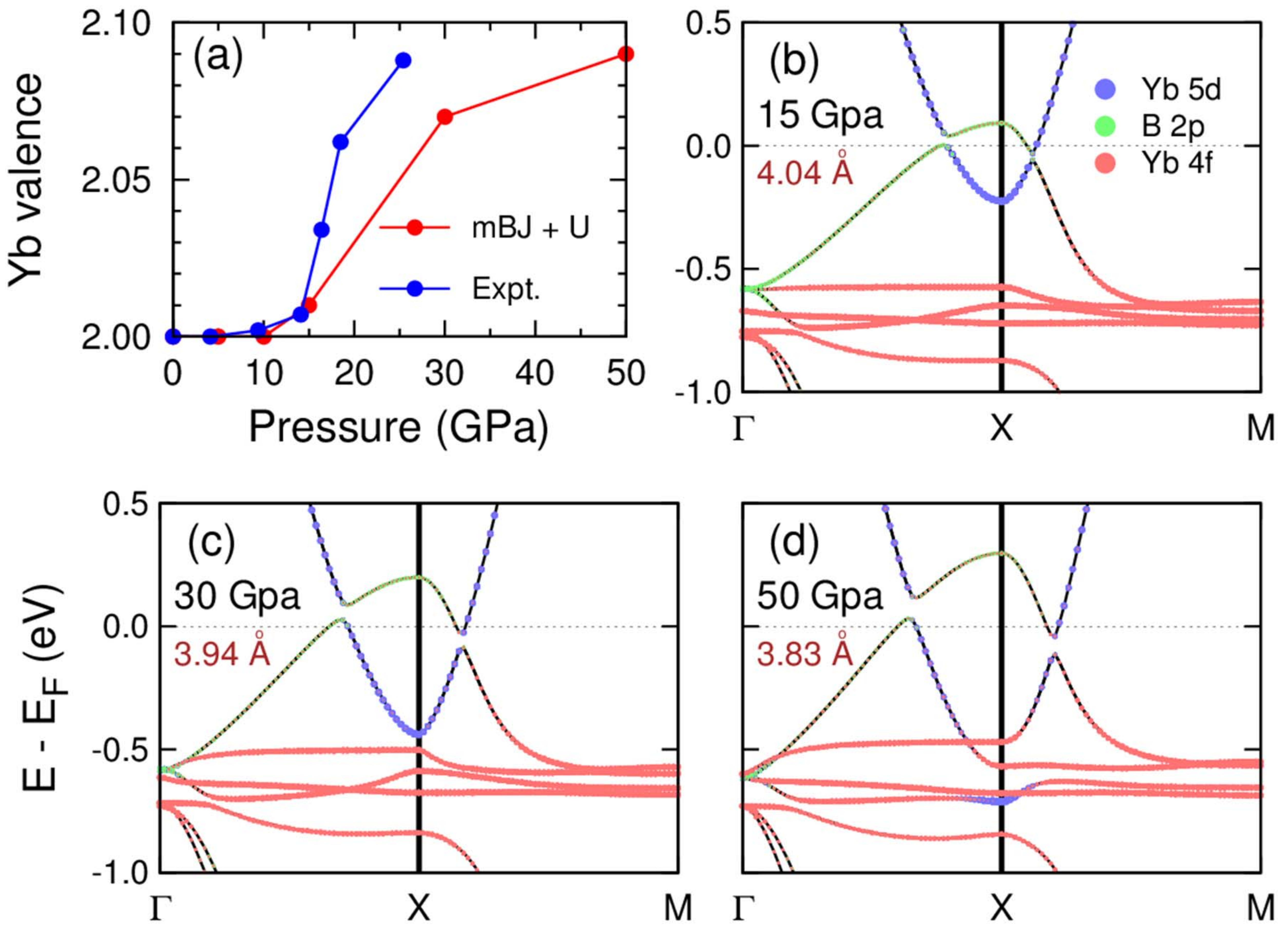}
\caption{(Color Online)
(a) Pressure-dependent Yb valence state.
For comparison, the experimental values are extracted from Ref. \cite{Zhou15}.
The mBJ + SOC + $U$ (7 eV) band structures (b) under P = 15 GPa,
(c) under P = 30 GPa, and (d) under P = 50 GPa.
}
\label{pressure}
\end{figure}

A recent pressure dependent study of YbB$_{6}$ \cite{Zhou15} observes key results of
(i) no structural transition up to 30 GPa from $x$-ray diffraction,
(ii) a rapid order-of-magnitude decrease in the resistivity up to 5 GPa,
(iii) a pressure region of rather constant resistivity
and Hall coefficient from 5 $\sim$ 15 GPa,
and (iv) a reemergence of thermally-activated resistivity
above 15 GPa accompanied by a small increase in Yb valency
from pure divalency to 2.09+.

The theoretical calculation at 15 GPa in Fig. \ref{pressure}(b)
shows a $p$-$d$ overlap band structure and indicates that
YbB$_{6}$ undergoes a semiconductor to semimetallic phase transition
at an intermediate pressure.
This occurs due to increase of $p$ and $d$ band widths
and their wave function overlap.
Such a $p$-$d$ gap to $p$-$d$ overlap transition naturally explains
the rapid initial decrease in resistivity with pressure,
also observed in early pressure-dependent transport of YbB$_{6}$ \cite{Sidorov91}.
A semimetallic state in the intermediate 5 $\sim$ 15 GPa pressure regime
is also suggested by the nearly constant Hall coefficient,
which is attributable to a balance between electron and hole carriers \cite{Zhou15}.
This transformation to semimetallic behavior under pressure provides
a further confirmation of the existence of a semiconductor gap at ambient pressure
where the ARPES experiments are performed.

The theoretical electronic structures for even greater pressures of 30 GPa and 50 GPa
in Figs. \ref{pressure}(c) and (d) show an increasing $p$-$d$ overlap such that
the Yb 4$d$ band ultimately touches the Yb 4$f$ band
which remains at nearly the same BE.
The Yb 4$f$ band exhibits only a small increase in bandwidth and slight
centroid shift to lower BE but still remaining at the BE larger than 0.5 eV.
Nevertheless there is an increased mixing of Yb 4$f$ character into the $p$-states,
as evidenced by the increasing band anti-crossing gapping
that results from the Yb 4$f$ SOC interaction.
The increasing Yb 4$f$ character above \EF\ implies
a decreased $f$-occupation and mixed-valency.
Quantitative analysis of the Yb valence under pressure is plotted in Fig. 4(a).
The resulting mixed-valence, less than 10\% at the highest pressure, compares favorably
to the experimental results derived from
Yb $L_{3}$ $x$-ray absorption measurements \cite{Zhou15}.
The experimental reemergence of a thermally activated resistivity ($dR/dT<0$) above 15 GPa
is plausibly due to the increasing 4$f$ SOC-induced local gapping,
whereas the overall resistivity rise due to gapping is weakened
due to the competition of the increasing $p$-$d$ overlap
and hence increasing hole and electron FS volumes.
The residual semimetallic conductivity can also explain
the observed experimental low temperature resistivity plateaus \cite{Zhou15}.


This theoretical investigation allows us to comment generally on
the feasibility of forming a TKI in actual materials.
Since $p$-$d$ states of opposite parity have inherently weak or negligible hybridization,
the topologically non-trivial band inversion will have difficulty
in forming a full insulator gap via hybridization alone.
Therefore some additional external influence is required to open up an insulating gap
of sufficient size to practically realize in-gap topological surface states.
Here for the example of YbB$_{6}$ under pressure,
the external influence is the hybridization mixing of the Yb 4$f$ states
with the $p$-states and its larger 4$f$ SOC-induced gapping.
However this effect is still too small for YbB$_{6}$
to develop a full BZ $p$-$d$ overlap gap at experimentally achievable pressures.

In conclusion, the flat-band conditions of the non-polar (110) surface
allow ARPES measurements to definitively show that YbB$_{6}$
is a non-Kondo non-TI semiconductor,
and it opens up a new method for the quantitative characterization of the bulk gap
of other divalent hexaborides.
This result is in good agreement with predictions of theoretical DFT+$U$ calculations
with proper treatment of 4$f$ correlations and inclusion
of well-established gap correction physics.
Only under pressure does the topologically non-trivial $p$-$d$ band inversion occur,
but the system still retains a semimetallic electronic structure
even up to high pressure beyond the onset of small Yb mixed valency.

Acknowledgments - This work was supported
by the Korean NRF (No.2011-0028736,
No.2013R1A1A2006416, No.2014R1A1A2056546, 2015R1A2A1A15053564),
Max-Plank POSTECH/KOREA Research Initiative (No. KR 2011-0031558),
the KISTI supercomputing center (No. KSC-2015-C3-007),
and the Deutsche Forschungsgemeinschaft via SFB 1170 (C06).
Experiments were supported by the U.S. DOE
at the Advanced Light Source (DE-AC02-05CH11231).

\newpage

\setcounter{table}{0}
\setcounter{figure}{0}

\begin{center}
{\bf \Large
{\it Supplemental Material:}\\
Electronic structure of YbB$_{6}$ : is it a Topological Insulator or not?
}
\end{center}

\def\EF{$E_\textrm{F}$}
\def\B6{B$_{6}$}
\def\cred{\color{red}}
\def\cblue{\color{blue}}

\renewcommand{\thefigure}{S\arabic{figure}}
\renewcommand{\thetable}{S\arabic{table}}

\section{ARPES of non-polar (110) surface}

\subsection{A. Yb divalency and wide EDC}

\begin{figure*}[t]
\includegraphics[width=13.5 cm]{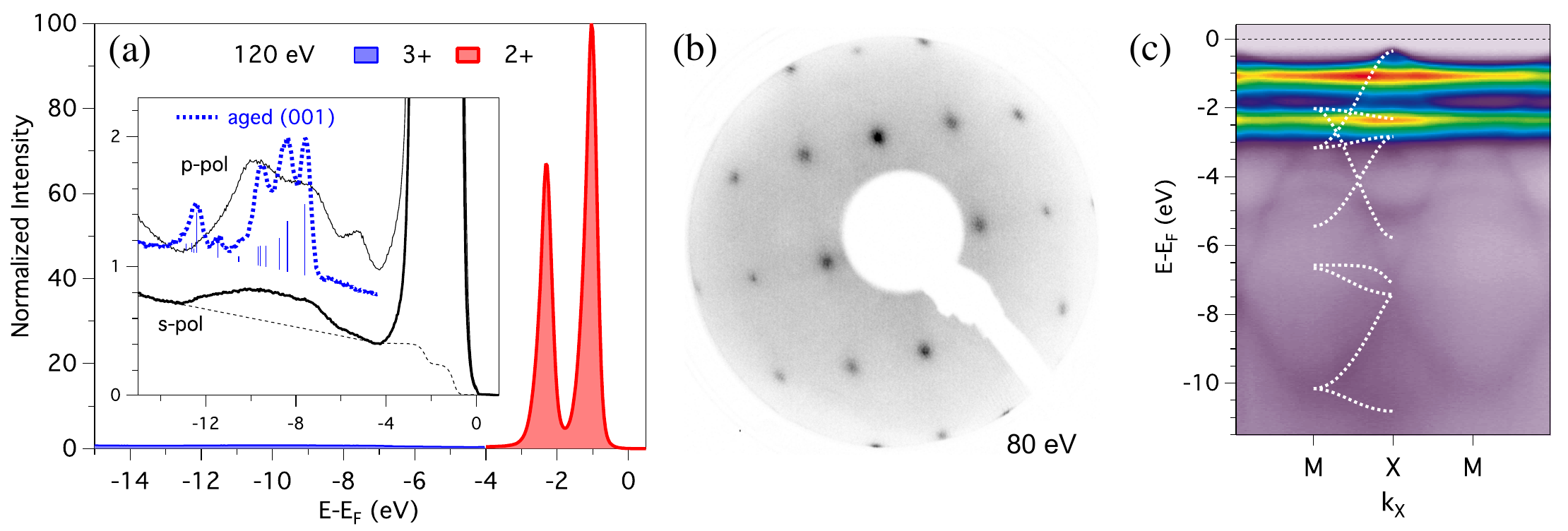}
\caption{
(a) Angle-integrated 120 eV spectra of Yb\B6\ (110) surface
demonstrating the pure divalency of Yb,
and ruling out the calculations in Fig. \ref{sdft1}(b,c).
(Inset) Magnified view of the overlapping B-$sp$ and Yb$^{3+}$ regions
showing the weaker B-$sp$ cross section for $s$-polarization
and the background subtraction for quantitative valence analysis.
Also shown (dashed line) for comparison is the signature spectral line shape
for Yb$^{3+}$ $f^{13} \rightarrow f^{12}$ multiplet peaks
acquired from an aged (001) cleaved surface of Yb\B6.
(b) Low-energy electron diffraction of the prepared (110) facet
showing 1 $\times$ 1 surface order.
(c) Wide-energy 120 eV X-point spectrum of the Yb\B6\ (110) surface
showing the divalent Yb $4f_{7/2}$ peak at 1.05 eV
and overplotted DFT bands along X-M.
}
\label{valence}
\end{figure*}

The deep binding energy (BE) of the bulk Yb 4\emph{f} levels
and lack of Yb$^{2+}$ spectral weight near the Fermi level \EF\
is strong experimental evidence for pure divalency in Yb\B6.
However, another important test for mixed-valency is the observation
and quantification of the relative amount of Yb$^{3+}$ spectral weight
in valence band photoemission measurements.
Previous XPS and UPS measurements \cite{sm-Kakizaki93,sm-Nanba93}
observed distinct Yb$^{3+}$ contributions to Yb 4\emph{p}
core-level and valence band spectra,
but the origin was ascribed to surface contributions.
Similarly, a weak ferromagnetic moment corresponding to 2\% bulk trivalency
was initially reported in Yb\B6\ \cite{sm-Gavilano03}.
But later measurements showed stoichiometric Yb\B6\ to be diamagnetic,
with surface ferromagnetism or paramagnetism appearing only
for non-stoichiometric samples \cite{sm-Kim07}.

In Fig. \ref{valence}(a), we demonstrate the pure divalency of Yb
from a wide valence band spectrum for a prepared Yb\B6\ (110)
surface measured at $h\nu$ = 120 eV.
The spectrum exhibits Yb$^{2+}$ $4f_{7/2}$ and $4f_{5/2}$ peaks
that dominate over the very weak 4 - 12 eV BE range
where Yb$^{3+}$ spectral weight would be found.
Overlapping in this energy range are B-\emph{sp} valence band states
whose relative cross section is low at this photon energy
and which can be further suppressed using \emph{s}-polarization
of the incident photons, as shown in the Fig. \ref{valence}(a) inset.

The line shape of the valence band structure,
shown in the expanded view of this 4 $\sim$ 12 eV range,
exhibits no evidence for any Yb$^{3+}$ spectral weight
whose distinct $f^{13} \rightarrow f^{12}$ final state multiplet peak structure
is illustrated in the inset of Fig. \ref{valence}(a) (dashed line).
This example of a Yb$^{3+}$ spectrum comes from a cleaved (001) surface
that initially looks like the pure divalent (110) surface spectra,
but then develops within only 4 hours the appearance of this Yb$^{3+}$ signature,
indicating the trivalent conversion of surface Yb atoms due to residual gas adsorption.
In contrast, the (110) surface with coplanar Yb and  \B6\  termination
and 1 $\times$ 1 surface order,  shown in Fig. \ref{valence}(b),
exhibits no spectral changes after greater than 12 hours of measurement.
Even if one integrates the B-\emph{sp} spectral weight
of the \emph{s}-polarization spectrum,
after employing the background subtraction shown in Fig. \ref{valence}(a),
and assigns it to Yb$^{3+}$,
one obtains an upper bound contribution of less than 1\%.
This tight experimental upper bound on the pure divalency of Yb in Yb\B6\
immediately rules out both of the theoretical predictions of $f$-$d$ \cite{sm-Weng14}
and $p$-$d$ \cite{sm-Neupane15,sm-Chang15} band inversion scenarios,
both of which predict mixed Yb valency.

A 12 eV wide angle-resolved spectrum measured at the 120 eV bulk X-point,
shown in Fig. \ref{valence}(c), also illustrates the wide B-\emph{sp} band dispersions
in the 4 - 12 eV range, and the absence of any $k$-independent Yb$^{3+}$ peaks.
The spectrum additionally illustrates the 1 eV and 2.3 eV binding energies
of the Yb $4f_{7/2}$ and Yb $4f_{5/2}$ states, respectively.
The Yb 4$f$ peaks are observed to be uniform across the prepared (110) sample surface,
in contrast to the reported energy shifts between cleaves,
and spatial inhomogeneity of the cleaved (001) surface \cite{sm-Xia14,sm-Xu14,sm-Frantzeskakis14}.
The boron hole band that forms the valence band maximum
is observed to disperse from below the Yb $4f_{5/2}$ state
and significantly hybridize with the \emph{f}-states
as it passes through towards \EF.
A small band width expansion of 1.05 of the DFT bands
was required to best match experiment.

\subsection{B. X-point location}

\begin{figure}[b]
\includegraphics[width=8.5 cm]{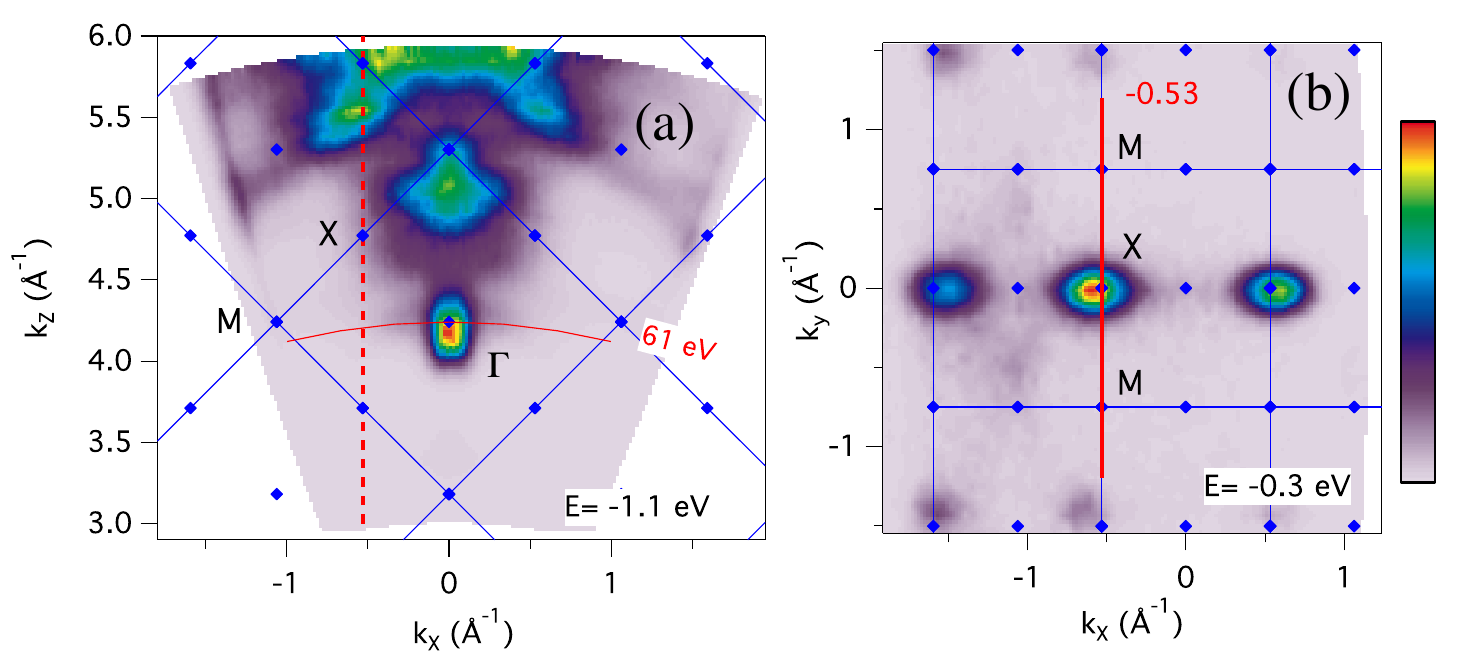}
\caption{
(a) Normal-emission photon-dependent map with the sample $(1\bar{1}0)$ direction
aligned to vertical spectrometer slit.
The identity of the $\Gamma$-point defines the inner potential ($V_{0}$ = 14 eV).
The vertical line indicates the off-normal plane of the photon-dependent maps
in Fig. 2(a) of the main text.
(b) Angle-dependent map at 120 eV, showing the $k$-location of the X-point spectrum
used for the band gap analysis Fig. 1 in the main text.
}
\label{xpoint}
\end{figure}

A suitable bulk X-point location for the band gap determination
was chosen based on photon-dependent and angular-dependent mapping of the momentum-space.
First a normal-emission photon-dependent $k_{x}-k_{z}$ map
was acquired with the sample $(1\bar{1}0)$ axis
aligned vertically with the spectrometer slit.
An energy cut at $-1.1$ eV, between the two 4\emph{f} states,
highlights a strong feature at 61 eV,
as shown in Fig. \ref{xpoint}(a) that aligns to the bulk Brillouin zone (BZ)
using an inner potential parameter of 14 eV.

Then the sample is rotated in azimuth to align the sample (100) axis
to the vertical spectrometer slit and an off-normal photon-dependent map
was acquired at fixed $k_{x} = -0.53 {\AA}^{-1}$
as indicated in Fig. \ref{xpoint}(a).
The $k_{y}-k_{z}$ map at fixed $k_{x}$,
presented in Fig. 2(a) of the main text,
with an energy cut 0.3 eV below \EF\
probing the valence band hole dispersion,
shows two strong hole-band features
at $k_{y}$ = 0 for $h\nu$ = 76 eV and 120 eV
as well as at four weaker intensity second BZ X-points
at $k_{y} = \pm1.5 {\AA}^{-1}$.

$h\nu$ = 120 eV was chosen for the $k_{x}-k_{y}$ angular-dependent map
in Fig. \ref{xpoint}(b) with the same sample orientation.
Similar to the photon-dependent map, strong valence band hole intensities
are observed at three different X-points at $k_{y}$ = 0 and also additionally
at four more X-points in the second BZ at $k_{y} = \pm1.5 {\AA}^{-1}$.
The bulk X-point at $k_{x} = -0.53 {\AA}^{-1}$ was used
for Fig. 1(e) in the main text
and was checked for consistency at other bulk X-points.

\subsection{C. K-dose}

\begin{figure}[b]
\includegraphics[width=8.5 cm]{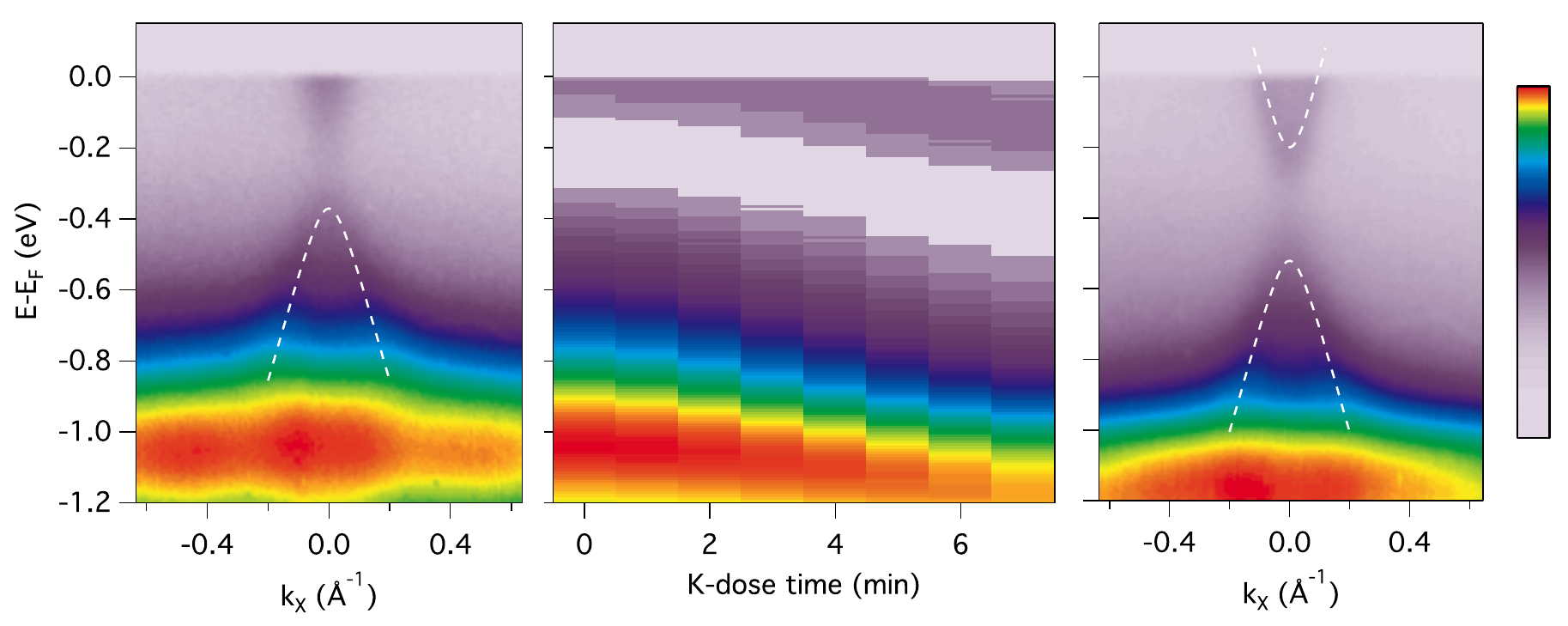}
\caption{
(a) X-point spectrum at $h\nu$ = 76 eV of the Yb\B6\ (110) surface before K-dosing
showing \EF\ pinned to the bottom of the Yb 5$d$ conduction band.
(b) K-dosing EDC image of the valence and conduction band energy shifts.
(c) X-point spectrum after K-dosing allowing
conduction band dispersion analysis for band gap determination.
}
\label{K-dose}
\end{figure}

To further characterize the conduction band minimum energy,
surface K-dosing was used to create a small $n$-type band bending
until the electron pocket was sufficiently deep
to observe its non-parabolic dispersion.
For this experiment, an X-point spectrum was obtained at 76 eV instead of 120 eV
due to the strong $d$-electron ($p$-hole) band intensity
for $s$ ($p$) polarization, allowing a
good quantitative analysis of the non-parabolic conduction (valence)
band dispersion and semiconductor gap evaluation
(see subsection D).

Fig. \ref{K-dose}(a) shows the X-point $p$-polarization spectrum before K-dosing.
Fig. \ref{K-dose}(b) shows a time/dosing dependent
energy distribution curve (EDC) cut at X
illustrating the energy shifts of the valence and conduction bands
as well as the Yb 4\emph{f} state.
The final EDC spectrum, shown in Fig. \ref{K-dose}(c),
is a sum of $p$- and $s$-polarizations
due to the weakening of the \emph{p}-hole band with K-dosing.
Because a rigid shift of both valence and conduction bands
is observed in Fig. \ref{K-dose}(b)
with no discernible surface band gap narrowing,
we can claim that this K-dosed analysis is reflective of a bulk gap
of magnitude 0.32 eV, based on non-parabolic dispersion visual fits to
the polarization-dependent data that best highlight either
the hole valence band or electron conduction band dispersions \cite{sm-Denlinger15}.

\subsection{D. Non-parabolic Dispersion}

A two-band $k\cdot p$ model for non-parabolic valence and conduction bands
in the presence of small band gaps \cite{sm-Kane57} was used to assist
in the determination of the valence band maximum and conduction band minimum energies.
The two band $k\cdot p$ correction to the free electron parabolic dispersion
can be written simply as \cite{sm-Lopez93}:
\begin{equation}
\varepsilon(k) = \frac{E_{g}}{2}\Bigg(\sqrt{1+\frac{4}{E_{g}}
	\cdot\frac{\hbar^{2}k^{2}}{2m^{*}}}-1\Bigg),
\label{kp}
\end{equation}
where $E_{g}$ and $m^{*}$ are the energy gap parameter
and the electron effective mass, respectively.

The energy gap dependence of Eq. (\ref{kp}) exhibits
the well-known progression towards a purely linear V-shaped dispersion at zero gap,
corresponding to a Dirac semimetal.
The valence band dispersion in Fig. 1(d) of the main text exhibits
linearity away from a distinctly rounded band maximum
and is nicely fit using the nominal experimental gap value of 0.3 eV
and an effective mass of $m^{*}$ = 0.12.
This dispersion shape for intermediate small gap values
contrasts sharply with topological Dirac cone surface states
in Bi$_{2}$X$_{3}$ (X = Se, Te), where the dispersion is mostly linear
near the Dirac point and deviates further away ({\it e.g.}, hexagonal warping).

The conduction band dispersion, revealed after K-dosing in Fig. \ref{K-dose}(c),
is similarly fit to $E_{g}$ = 0.3 eV and $m^{*}$ = 0.1.

\section{Computational details}

The band calculations were performed by using
the full-potential linearized augmented plane-wave (FLAPW)
band method, as implemented in the WIEN2K package \cite{sm-wien2k}.
For the DFT calculations, the PBE (Perdew-Burke-Ernzerhof) exchange-correlation
functional was used in the GGA (generalized-gradient approximation).
The spin-orbit coupling (SOC) was included in a way of the second variational method.
In the GGA + SOC + $U$ method, a correlation energy of $U=7$ eV was
chosen to obtain the correct experimental value of the
Yb 4\emph{f} BE of $\approx$ 1 eV,
which is consistent with the previous calculations \cite{sm-Jun07,sm-Kunes04}.
The mBJ potential corresponds to an orbital-independent
semilocal exchange potential
that mimics the behavior of orbital-dependent potentials
and has been shown to provide band gap corrections
in good agreement with the improved many-body
but more computation-demanding GW calculation \cite{sm-TB09,sm-Singh10}.
A single pass open-core pseudopotential GW calculation
was also performed using VASP \cite{sm-vasp},
which effectively removes the Yb 4\emph{f} levels from the calculation.
This is reasonable in view of the deep 4\emph{f} BE in Yb\B6.

In the GGA + $U$ ($U$: on-site Coulomb interaction of Yb $4f$ electrons) calculations,
the scheme of a fully-localized limit was used for the 4\emph{f} electrons.
The BZ integration was done with a 17 $\times$ 17 $\times$ 17 $k$-mesh, and,
in the FLAPW calculations, muffin-tin radii
$R_{MT}$'s of Yb and B were chosen to be 2.50 and 1.59 a.u.,
respectively, and the plane-wave cutoff was $R_{MT}K_{max}$ = 7.

For the GW calculation, a single pass open-core pseudopotential
G$_{0}$W$_{0}$ calculation was performed using VASP
with a 4 $\times$ 4 $\times$ 4 $k$-mesh,
400 unoccupied bands and 80 mesh size in the omega frequency.

YbB$_{6}$ crystallizes in the CsCl-type structure, like SmB$_{6}$.
The ambient pressure calculations were performed
using the experimental YbB$_{6}$ lattice parameter of \emph{a} = 4.1792 {\AA}
and internal crystal parameter $x_{B}$ = 0.202
that specifies the relative inter-octahedron B-B bond length \cite{sm-Jun07}.
For pressure-dependent calculations,
a linear relation between the lattice constant and applied pressure up to 30 GPa
was established using $x$-ray diffraction data in Ref. \cite{sm-Zhou15},
before starting the $x_{B}$ relaxation.
To obtain the structure information under 50 GPa,
where the lattice constant was not accessible in Ref. \cite{sm-Zhou15},
the fully relaxed calculation was performed.
All electronic structure calculations under pressure
were done using the mBJ + SOC + $U$ (7 eV) scheme.
We have checked the pressure dependent $U$ value
by a constrained LDA calculation \cite{sm-Madsen05}
and found that $U$ $\approx$ 7 eV is obtained over the whole pressure range,
which could be interpreted as showing the strong localization of the Yb $4f$ orbital.

\begin{figure}[t]
\includegraphics[width=8.5 cm]{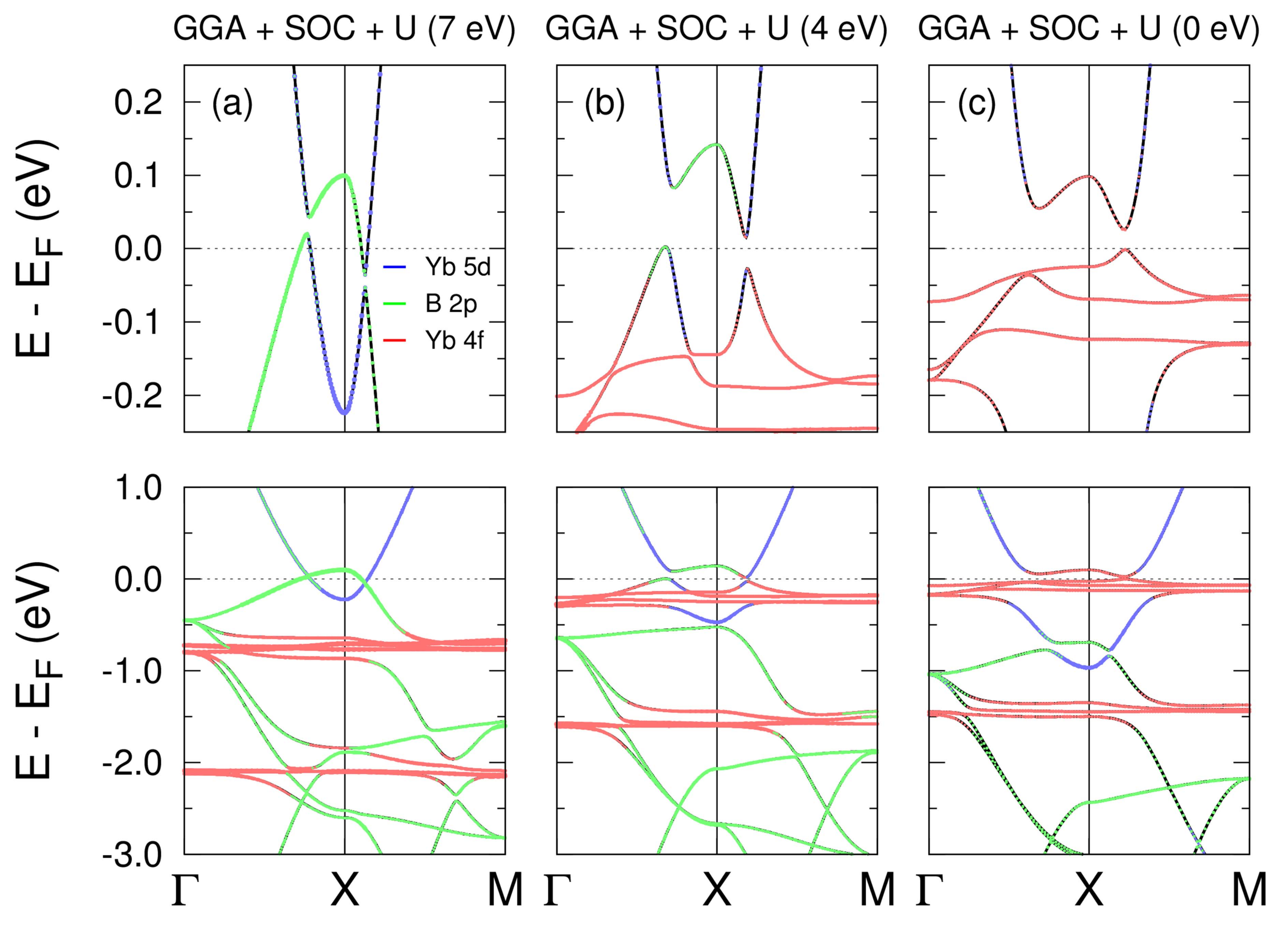}
\caption{
DFT-Wien2k band structures of Yb\B6.
(a) GGA + SOC + $U$ (7 eV) calculation yields a semimetallic
$p$-$d$ overlap with anticrossing gaps.
(b) GGA + SOC + $U$ (4 eV) calculation yields SOC enhanced
gapping of overlapping $p$-$d$ states,
similar to the VASP calculation by Neupane \emph{et al.} \cite{sm-Neupane15}.
(c) GGA + SOC + $U$ (0 eV) calculation yields $f$-$d$ band inversion,
similar to the LDA + Gutzwiller calculation by Weng \emph{et al.} \cite{sm-Weng14}.
}
\label{sdft1}
\end{figure}

\section{DFT + SOC + $U$ band structures for Y\MakeLowercase{b}\B6\ with varying $U$}

\begin{figure*}[t]
\includegraphics[width=15.5 cm]{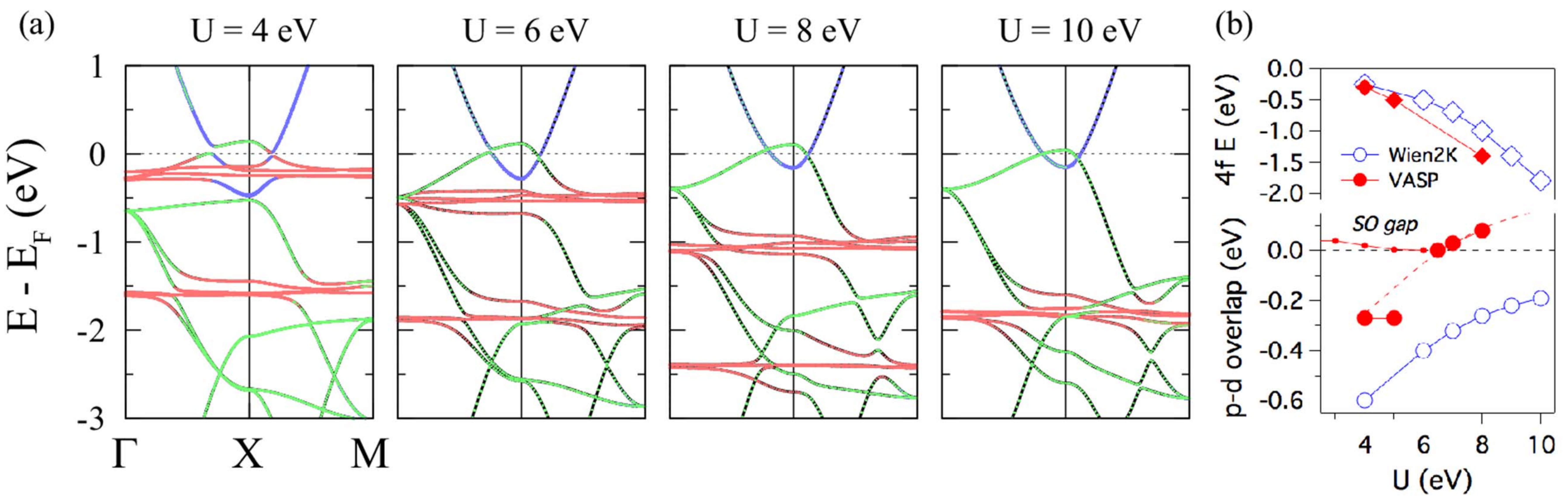}
\caption{
(a) The GGA + SOC + $U$ calculations using Wien2k,
showing the persistence of the \emph{p}-\emph{d} semimetal overlap
up to unphysically high values of $U$.
Red, blue, and green color symbols represent
Yb 4\emph{f}, Yb 5\emph{d}, and B 2\emph{p} characters, respectively.
(b) Summary comparison of the \emph{p}-\emph{d} band overlap
and the binding energy of Yb 4\emph{f} band ($E_{4f}$)
between our Wien2k and the VASP calculations by Chang \emph{et al.}
\cite{sm-Neupane15,sm-Chang15}.
The SOC-induced \emph{p}-\emph{d} band gap by Chang \emph{et al.} (SO gap) are
also plotted.
}
\label{sdft2}
\end{figure*}

Two recent calculations predict YbB$_{6}$
to be a topological insulator (TI) based on $f$-$d$ band inversion \cite{sm-Weng14}
or $p$-$d$ inversion \cite{sm-Neupane15,sm-Chang15}.
The first was an LDA + Gutzwiller calculation with $U$ = 6 eV
that erroneously predicted a topological $f$-$d$ inverted band structure
very similar to Sm\B6\ \cite{sm-Lu13},
with a highly incorrect 4$f$ BE of $\approx$ 0.1 eV.
The LDA + Gutzwiller result \cite{sm-Weng14} is empirically very similar
to a $U$ = 0 DFT result in Fig. \ref{sdft1}(c) in terms of the 4$f$ BE
and the $f$-$d$ band inversion at \EF.

A second proposed topological band structure for Yb\B6,
based on DFT + $U$ pseudopotential VASP calculations
using small $U$ values of 4 eV \cite{sm-Neupane15} or 5 eV \cite{sm-Chang15},
obtains a $p$-$d$ band overlap
but with a small insulating gap throughout the bulk BZ,
similar to that in Fig. \ref{sdft1}(b) \cite{sm-Replica}.
The effect of the smaller $U$ values is a smaller Yb 4$f$ BE,
0.3 (0.5) eV for $U$ = 4 (5) eV, whose closer proximity to
the $p$-$d$ overlap region at \EF\
results in significant hybridization of the Yb 4$f$ and B 2$p$ states.
This allows the Yb 4$f$ SOC to open a larger $p$-$d$ gap
that otherwise is intrinsically tiny
due to the smallness of the 5$d$ SOC.
The hybridization of Yb 4$f$ states into the $p$-$d$ overlap region
also results in a mixed-valent Yb 4$f$-occupation evaluated
to be $n_{f}$ $\approx$ 13.8,
which does not agree with the experimental divalent state of Yb.
Chang \emph{et al.} \cite{sm-Chang15} also reported that
the SOC-induced gapping of the $p$-$d$ overlap at $U$ = 4 $\sim$ 5 eV
evolves into a topologically trivial $p$-$d$ gap above $U$ = 6.5 eV.
This result is in contradiction to the earlier literature result \cite{sm-Jun07}
and our Wien2k result shown in Fig. \ref{sdft1}(a).
As shown in Fig. \ref{sdft2}, the DFT + $U$ calculations
for increasing $U$ yield only a small gradual decrease
in the $p$-$d$ overlap energy
without opening up a semiconducting $p$-$d$ gap
even for unphysically large $U$ values as high as 15 eV.
This is consistent with a decreasing influence of the Yb 4$f$ states
on the $p$-$d$ states as they are pushed farther away from \EF.
Namely, large $U$ is not the formative parameter for the semiconductor gap.


One might consider that the $U$ $>$ 6.5 eV VASP prediction
of a topologically trivial semiconductor
is in agreement with the Wien2k mBJ + SOC + $U$ semiconductor result in Fig. 3(b)
of main text.
However this is entirely fortuitous because the VASP calculation does not contain
the band gap renormalization treatment of the mBJ or GW calculations in Fig. 3(b).
Chang \emph{et al.} \cite{sm-Chang15} also presented a GW calculation for Yb\B6\
but for $U$ = 0 with the goal of demonstrating that GW alone
cannot produced the $p$-$d$ band inversion.
In contrast, the VASP GW calculation in Fig. 3(b) is an open-core calculation,
which simulates a large effective $U$ that removes the Yb 4$f$ states far from \EF.
Thus the GW calculation in Fig. 3(b) treats
the $p$ and $d$ correlations that renormalize the $p$-$d$ overlap
into a small semiconductor gap.

\begin{figure}[b]
\includegraphics[width=8.5 cm]{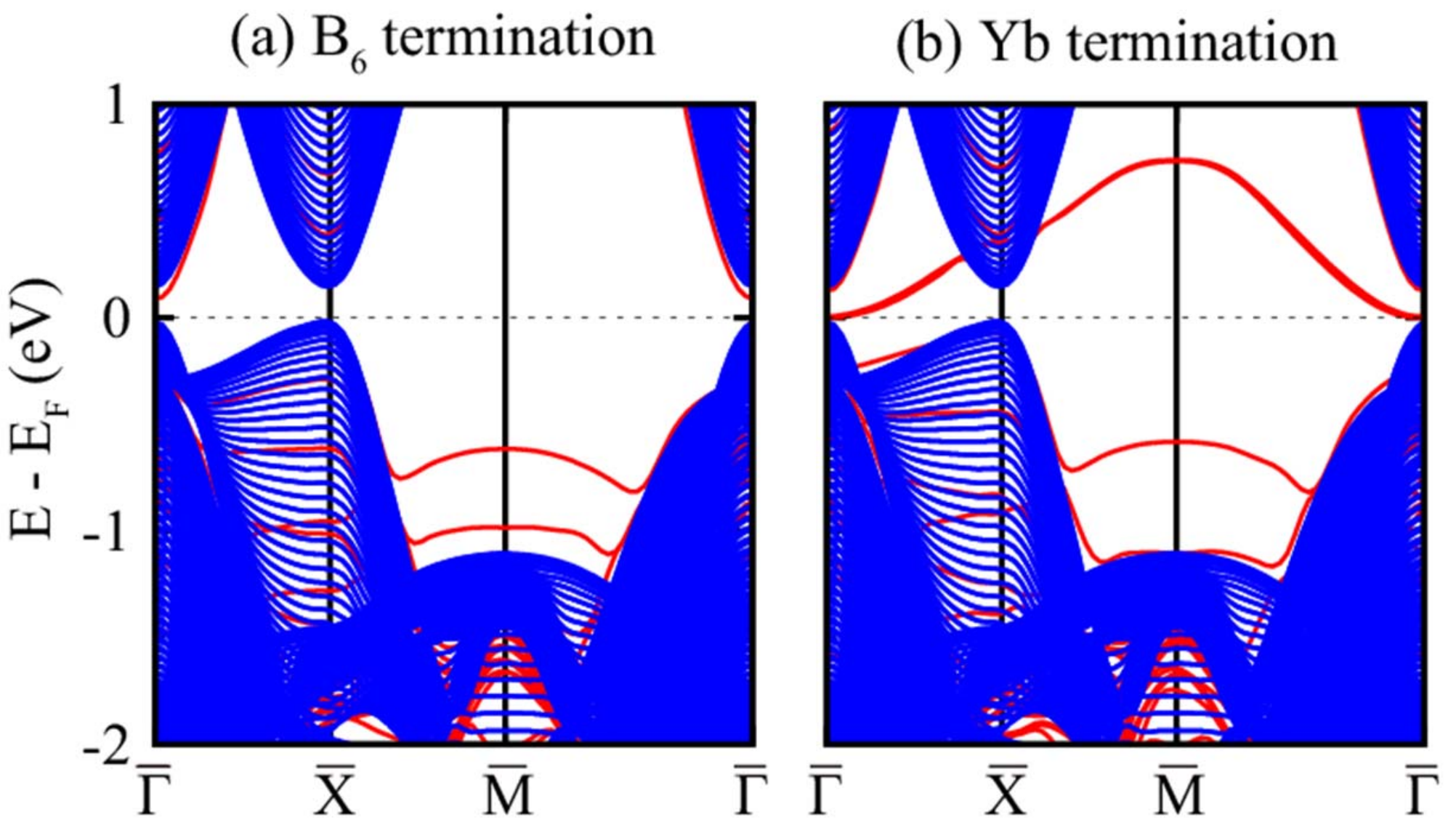}
\caption{
DFT-VASP open-core slab calculations for Yb\B6\ (001) surface.
For the purpose of getting a trivial \emph{p}-\emph{d} gap without the mBJ scheme,
a lattice constant of Yb\B6\ is enlarged to be $a = 4.5 {\AA}$.
(a) \B6\ termination.
(b) Yb termination.
Both of terminations give topologically trivial surface states.
The surface states (red lines) below \EF\ are mostly of B-$p$ character,
while those above \EF\ in (b) have mixed B $p$ and Yb $d$ characters.
}
\label{slab}
\end{figure}

We have also checked DFT + $U$ electronic structures using VASP.
The crystal structure was fully relaxed using the GGA + SOC scheme
($a$ = 4.13 {\AA}, $x_{B}$ = 0.201).
The resultant crystal structure was quite similar to the experimental structure.
Band structures of Yb\B6\ were calculated with various $U$ values
in both relaxed and experimental structures,
and it was found that the band structures for the two cases are almost the same.
We have found that our VASP results are quite consistent with our Wien2k results.
There is a sensitivity of the $p$-$d$ overlap to
the internal structural parameter $x_{B}$ \cite{sm-Lee05}.
One could speculate that differences in structural input could be
a possible source of the discrepancy between our Wien2k (and VASP)
and the published VASP calculations \cite{sm-Neupane15,sm-Chang15},
but no structural information (lattice constant or $x_{B}$)
was provided in Refs. \cite{sm-Neupane15,sm-Chang15}.

Figure \ref{sdft2}(b) shows the comparison between the VASP results
by Chang \emph{et al.} \cite{sm-Chang15} vs. our Wien2k results,
which highlights the discrepancy for $U$ $>$ 6 eV.
We can note discrepancies between two in
(i) the binding energy of Yb 4$f$ band ($E_{4f}$)
and (ii) $p$-$d$ overlap.
The VASP results of the latter by Chang \emph{et al.}
look merely shifted Wien2k results.

In summary, we note  that the incorrect $f$-$d$ overlap \cite{sm-Weng14},
and $p$-$d$ overlap \cite{sm-Neupane15,sm-Chang15} calculations
did not cite the earlier Wien2k calculation of Ref. \cite{sm-Jun07},
and hence failed to compare and discuss their discrepancies with the published literature.
We also note that the complicated intermixing of all three $p$, $d$ and $f$ states
near \EF\ with interplay of 4$f$ SOC and mixed valency discussed in the main text
for the high-pressure phase of Yb\B6\  is the same physics at the heart of
the erroneous claim of a TI $p$-$d$ overlap gap at ambient pressure \cite{sm-Neupane15, sm-Chang15},
albeit errantly realized from a too small value of $U$.

A remaining open question about the ambient pressure phase
is the quantitative disagreement between the experimental 0.3 eV and theoretical 0.1 eV gaps,
which would be worthwhile to be investigated theoretically.
Further theoretical investigation of the high pressure phase
is warranted to address whether any low energy scale dynamical correlations
({\it e.g.} Kondo effect) can emerge from the small mixed valency,
or whether any non-trivial topological $p$-$d$ parity inversion effects
can be realized while in a partially gapped semimetallic state.

\section{DFT slab calculation for Y\MakeLowercase{b}\B6\ (001) surface}

In order to examine the surface states,
we have also carried out DFT open-core slab calculations for the Yb\B6\ (001) surface
using VASP.
We have used slab structures composed of 9 Yb and 8 \B6\ layers for Yb termination
and 8 Yb and 9 \B6\ layers for \B6\ termination,
and for both terminations, $\approx$ 30 {\AA} vacuum layers were considered.
We have first relaxed the slab structures for both terminations,
and then self-consistent calculations were performed
with a 10 $\times$ 10 $\times$ 1 $k$-mesh.
For the purpose of getting a topologically trivial $p$-$d$ bulk gap in the
DFT + SOC + $U$ (7 eV) without employing mBJ,
we employed an increased lattice constant of $a = 4.5 {\AA}$
in the slab calculations.

As shown in Fig. \ref{slab}, surface states (red lines) appear
for both \B6\ and Yb terminations.
These surface states, however, are just topologically trivial surface states,
which are often observed in the hexaboride systems \cite{sm-Zhu13,sm-Monnier04}.
The Dirac-cone-like surface states observed in ARPES are not obtained.
In fact, such Dirac cones are expected to be quantum well states
coming from the band bendings in the polar (001) surfaces with different
charge terminations \cite{sm-Frantzeskakis14,sm-Denlinger15}.

The DFT open-core slab calculations for the non-polar (110) surface
also produce similar surface states, one deep surface band of B $p$ character
and two surface bands near \EF\ of mixed B $p$ and Yb $d$ characters
that arise from broken bond states from $two$ surface B atoms  \cite{sm-CJKang15}.
All of those surface states are topologically trivial.

\section{Link between Y\MakeLowercase{b}\B6\ and S\MakeLowercase{m}\B6\ }

There are strong differences in issues of the two materials Yb\B6\ and Sm\B6\,
{\it i.e.}, pure-divalency and $p$-$d$ overlap/semiconductor gap issues in Yb\B6\
and mixed valency and $f$-$d$ hybridization gap issues in Sm\B6.
Also note that while Yb\B6\ (001) ARPES shows ``false" in-gap states
that are actually band-bending quantum well states,
the Yb\B6\ (110) ARPES show no states within the 0.3 eV semiconductor gap,
thereby disproving the TI scenario for Yb\B6.

In contrast for Sm\B6\ the 20 meV hybridization gap exhibits similar
ARPES in-gap states for both the polar (001)
and non-polar (110) surfaces \cite{sm-Denlinger16}.
The existence of the X-point in-gap states for Sm\B6\
is insensitive to the different polar surface terminations,
and similar in-gap states are observed
for the non-polar (110) surface too \cite{sm-Denlinger16}.
Hence the polar nature alone cannot explain the in-gap states,
although polar effects are still being discussed for the (001) surface,
{\it e.g.}, in the context of 2D-like $d$-states outside the gap \cite{sm-Hlawenka15}.

We have also checked the mBJ scheme for Sm\B6.
The reason why we used the mBJ scheme for Yb\B6\ is to consider
the correlation effects of delocalized Yb 5$d$ and B 2$p$ electrons near \EF.
While it is very crucial for Yb\B6 that the mBJ correction
changes the $p$-$d$ overlap into a $p$-$d$ trivial gap,
for Sm\B6, the Sm 4$f$ bands are dominating near \EF,
and so the mBJ method does not play a crucial role.
In fact, we have confirmed for Sm\B6 that the DFT and mBJ results are essentially
the same \cite{sm-CJKang15}.

For Sm\B6, dynamical mean-field theory (DMFT) is essential
to address the dynamic energy correlations
that result in strong low energy scale renormalization of the $f$-states
near \EF\ as well as the mixed valency.
In contrast, for Yb\B6, with the large binding energy of the 4$f$ states and pure divalency,
one would not expect any low energy scale dynamical correlations
to emerge from DMFT calculations.
Indeed we have checked and confirmed for Yb\B6 that DFT + DMFT gives
essentially identical results as DFT + $U$,
{\it i.e.}, a proper 4$f$ binding energy and an uncorrected $p$-$d$ overlap \cite{sm-CJKang15}.
The latter is because the correlation effects of delocalized $p$ and $d$ electrons are
not treated within DMFT.

\end{document}